\documentclass[twocolumn,showpacs,amsmath,amssymb]{revtex4}
\usepackage{graphicx}

\begin{document}

\title
{\Large \bf
 Graphene-based modulation-doped superlattice structures
}

\author{\bf
Dima Bolmatov$^{a,b}$\footnote{e-mail: d.bolmatov@qmul.ac.uk} and Chung-Yu Mou$^{b,c}$}

\affiliation{$^{a}$ Department of Physics, Queen Mary University of London, Mile End Road, London, E1 4NS, UK \\
$^{b}$ Department of Physics, National Tsing Hua University, Hsinchu 300 Taiwan \\
$^{c}$ National Center for Theoretical Sciences, Hsinchu 300, Taiwan
}


\begin{abstract}
The electronic transport properties of graphene-based superlattice structures are investigated. A graphene-based  modulation-doped superlattice structure geometry is proposed and consist of periodically arranged alternate layers: InAs/graphene/GaAs/graphene/GaSb. Undoped graphene/GaAs/graphene structure displays relatively  high conductance and enhanced mobilities at elevated temperatures unlike modulation-doped superlattice structure more steady and less sensitive
to temperature  and robust electrical tunable control on the screening length scale. Thermionic current density exhibits enhanced behaviour due to presence of metallic (graphene) mono-layers in superlattice structure. The proposed superlattice structure might become of great use for new types of wide-band energy gap quantum devices.

\end{abstract}
\maketitle

\section{Introduction}
Graphene is a single layer of carbon atoms densely packed in a honeycomb structure was recently first isolated in its free-standing form \cite{Nov-1,Zhang-1}. However, its unusual material and
physical properties have already captured the interest of many researchers working
in condensed-matter physics \cite{Chris-1,Mou-1,Fal-1,Sevin-2}. This two-dimensional material is of very high quality,
extremely strong, exhibits ballistic electronic transport on the micrometer scale at room
temperature, can be chemically doped and its conductivity can be controlled with an
electric field \cite{Sevin-1,HHLin-1,Li-1,Ost-1}. Graphene has a linear gapless spectrum, and therefore exhibits metallic conductivity even in the limit of nominally zero carrier concentration \cite{Ben-1,Bol-1,Stroscio-1}. At the same time, most electronic applications rely on the presence of a gap between the
valence and conduction bands \cite{Lozovik-1,Agr-1,Jan-1,Hong-1,Esm-1,Fal-2,Ber-1}.

The continuing enhancing of quantum electronics devices poses new challenges to the semiconductor industry for each new device generation \cite{Ali-1,Daw-1,Shih-1}. At the mesoscopic scale there are important quantum effects and the materials which were working well in previous device generations do not perform properly at the nanoscale and new materials need to be introduced \cite{Kwo-1}. Eventually, not only the materials \cite{Kostya-1} but also the basic device operation principles \cite{Feig-1} and geometries need to be revised \cite{Bax-1}.

Superlattices have been used to filter the energy of electrons \cite{Sev-1,Abe-1}. The band structure can be tuned by varying the composition and thickness of the layers \cite{Sor-1}. In fact superlattices are widely used in an applications that has nothing to do with their electronic properties \cite{Yu-1,Niu-1,Ou-1}. This is to improve the cleanliness of material during growth \cite{New-1}. The structures which will be discussed below are vertical, in the sense that current flows along the direction of growth or normal to the interfaces \cite{Sato-1}. The obvious way of introducing carriers, used in classical devices, is to dope the regions where electrons or holes desired \cite{Guin-1,Ndu-1}. The solution is remote or modulation doping, where the doping is grown in one region but the carriers subsequently migrate to another \cite{Nori-1}. Thus modulation doping has achieved two benefits: it has separated electrons from their donors (holes from their acceptors) to reduce scattering by ionized impurities, and the electrons (holes) to two dimensions.

In this work we propose a new graphene-based  modulation-doped superlattice structure geometry which consist of periodically arranged alternate layers: InAs/graphene/GaAs/graphene/GaSb. In Fig.1 is illustrated graphene device tunneling structure: monolayers of graphene sandwiched between thin layer of InAs in the bottom of the quantum device and thin layer of GaSb in the top. In the middle of proposed superlattice structure is placed gallium arsenide (GaAs) which has a higher saturated electron velocity and higher electron mobility and  has some electronic properties which are superior to those of silicon. GaAs devices are relatively insensitive to heat, generate less noise than silicon devices when operated at high frequencies. GaAs layer is captured by two graphene mono-layers. Weak anti-localization,  mobility and carrier density of the graphene allow us to consider this geometry as $intrinsic$ semiconductor structure which we treat in the first section of our work. The weak van der Waals forces that provide the cohesion of multilayer graphene stacks do not always affect the electronic properties of the individual graphene layers in the stack. 

Our idea here is to create two high-conductivity channels in the current-spreading graphene layers, one of which is sandwiched by GaAs (gapped material) in the top and by InAs (electrons-doped region) in the bottom, another is sandwiched by GaAs (gapped material)  in the bottom and by GaSb (holes-doped region) in the top. Electronic transport properties of this structure are investigated in the second part of current issue.
\begin{figure}
	\centering
\includegraphics[scale=0.35]{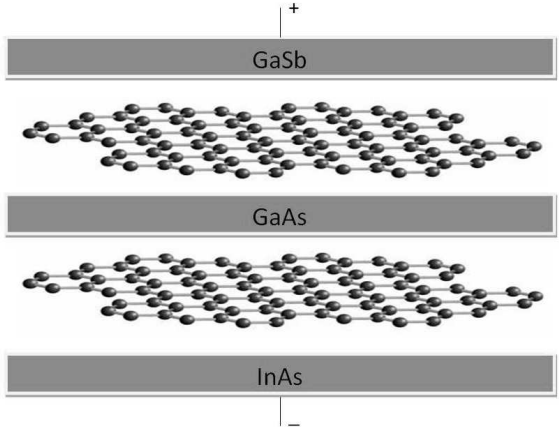} 
\caption{Graphene device tunneling structure. Two monolayers of graphene sandwiched between thin layer of InAs in the bottom of the quantum device and thin layer of GaSb in the top. In the middle of proposed superlattice structure is placed gallium arsenide (GaAs).}
	\label{fig-1}
\end{figure}
\section{electronic properties}
Pure semiconductors, which are free from impurities, are called $intrinsic$ $semiconductors$. Ideally, the intrinsic conductivity is zero at 0 K and increases with temperature owing the thermal excitation of electrons from the valence to the conduction band. The electrons thus excited, leave holes in the valence band. The electrical conductivity in these materials will be form both the electrons in the conduction band and the holes in the valence band. The conductivity in these materials can be written as
\begin{eqnarray}
\sigma=n_{e}e\mu_{e}+n_{h}e\mu_{h}
\end{eqnarray}
where $n_{e}$ and $n_{h}$ are the electron and hole concentrations, $\mu_{e}$ and $\mu_{h}$ are the electron and hole mobilities (in graphene $200,000 \ cm^2/ V\cdot s$, in two-dimensional electron gases $3,000,000 cm^2/\ V\cdot s$) respectively, and $e$ the charge of electron or hole. The number of electrons available in the conduction band depends on two factors, viz. the number of electronic energy levels available in the conduction band and the extent to which these energy states are occupied. First is given by the density of states and second factor comes from the Fermi-Dirac distribution function. 

If $f(E)$ gives the probability of finding the electron in the energy state with energy $E$, then, $1-f(E)$ will give the probability for the electron not being found in that state or the hole probability as the total probability is 1. The total number $N$ of electron is given by
\begin{eqnarray}\label{number}
N=\sum_{i}\frac{1}{e^{\beta(E_{i}-\mu)}+1}+\sum_{j}\frac{1}{e^{\beta(E_{j}-\mu)}+1}
\end{eqnarray}
where $E_{i}$ is energy level in the conduction band and $E_{j}$ is an energy level in the filled band. In the case of an intrinsic semiconductor the total number of electron states in the full band is equal to $N$, that is $\sum_{j}1=N$. Hence one has, from Eq.(\ref{number})
\begin{eqnarray}
\nonumber
\sum_{i}\frac{1}{e^{\beta(E_{i}-\mu)}+1} &=&  \sum_{j} \left( 1 - \frac{1}{e^{\beta(E_{j}-\mu)}+1}\right) \\
 &=& \sum_{j}\frac{1}{e^{\beta(-E_{i}+\mu)}+1}
\end{eqnarray}
This equation shows the equality of the number of conduction electrons (the left hand side) and number of holes in the filled band (the right hand side) in intrinsic semiconductors, that is 
\begin{eqnarray}\label{cons}
n_{e}=n_{h}
\end{eqnarray}
The number of electrons in the conduction (holes in the valence) band is obtained by integration the following expression $n_{e}=\int^{\infty}_{E_{g}}D(E)f(E)dE$ ($n_{h}=\int^{0}_{-\infty}D(E)(1-f(E))dE$). Here the origin of energy is taken at the top of the filled band and the assumption $E_{g}\gg kT$ is made. Substituting for $D(E)$ and $f(E)$ in the above, we then have for electrons
\begin{eqnarray}
n_{e}=\frac{1}{\pi}\int_{E_{g}}^{\infty} \frac{E-E_{g}}{(\hbar \upsilon_{F}^{e})^{2}}e^{(\mu-E)/kT}dE
\end{eqnarray}
and for holes
\begin{eqnarray}
n_{h}=\frac{1}{\pi}\int_{-\infty}^{0}\frac{(-E)}{(\hbar\upsilon_{F}^{h})^{2}}e^{(-\mu+E)/kT}dE
\end{eqnarray}
Now the electron and hole densities become
\begin{eqnarray}\label{densitya}
n_{e} &=& \frac{1}{\pi}\left(\frac{kT}{ \hbar \upsilon_{F}^{e}}\right)^{2}e^{(\mu-E_{g})/kT}
\end{eqnarray}
\begin{eqnarray}\label{densityb}
n_{h} &=& \frac{1}{\pi}\left(\frac{kT}{ \hbar \upsilon_{F}^{h}}\right)^{2}e^{-\mu/kT}
\end{eqnarray}
From equations (\ref{densitya}), (\ref{densityb}) and (\ref{cons}) one can determine $e^{\mu/kT}$ as
\begin{eqnarray}\label{mu}
e^{\mu/kT}= \frac{\upsilon^{e}_{F}}{\upsilon^{h}_{F}}e^{E_{g}/2kT}
\end{eqnarray} 
Hence one has from (\ref{densitya}) and (\ref{densityb})
\begin{eqnarray}
n_{e}=n_{h}=\frac{1}{\pi\upsilon^{e}_{F}\upsilon^{h}_{F}}\left(\frac{kT}{\hbar}\right)^{2}e^{-E_{g}/2kT}
\end{eqnarray}
From equation (\ref{mu}) one yields 
\begin{eqnarray}\label{potential}
\mu=\frac{1}{2}E_{g}+kT\log{\frac{\upsilon_{F}^{e}}{\upsilon^{h}_{F}}} 
\end{eqnarray}
The chemical potential $\mu$ in (\ref{potential}) lies in the vicinity of the middle of the forbidden energy gap provided that the value of $\log{\frac{\upsilon_{F}^{e}}{\upsilon^{h}_{F}}}$ ($\upsilon_{F}^{e}\simeq 1.11\times10^6 \ m/s $, $\upsilon^{h}_{F}\simeq 1.04×10^6 \ m/s$ in graphene mono-layer and $\upsilon^{e}_{F}\simeq 1.10×10^6 \ m/s$, $\upsilon^{h}_{F}\simeq 1.07×10^6 \ m/s$ in layered graphene correspondently) is of the order of the unity and that the temperature is well below the value of $E_{g}/k$. Hence at ordinary temperature the relations $E_{g}/k\gg T$, $E_{g}-\mu\gg kT$ and $\mu\gg kT$ are satisfied.
\section{Modulation-doped superlattices}
\subsection{Neutrality}
Because semiconductors contain mobile electric charges, they tend to be electrically neutral, which is to say they contain equal amounts of positive and negative charge. It is interesting to see how large a region can be non-neutral, without there being large potential differences.

In this work we focus on consideration of graphene-based modulation-doped superlattice structures. Inherently InAs and GaSb are doping layers in superlattice structure, graphene mono-layers make electric carriers highly mobile and GaAs layer is active layer where we propose a varying potential tends to a constant value, taken as zero. In the constant potential, the hole (electron) density $n_{0}$ will equal the acceptor (donor) density $A$ (below not loosing the generality we will regard the p-type (hole) carriers, where the potential has charged to $V$, the hole density will be controlled by the Maxwell-Boltzman relation: $n=n_{0}\exp{(-eV/kT)}$, the assumption $E_{g}\gg kT$ is made. Poisson's equation for this case is
\begin{eqnarray}\label{P}
\frac{d^2 V}{dx^2}=\frac{-e}{\epsilon_{s}\epsilon_{0}}(n-A)=\frac{-\epsilon n_{0}}{\epsilon_{s}\epsilon_{0}}(\exp{(-eV/kT)}-1)
\end{eqnarray}
where $\epsilon_{0}$ is the permittivity, and $\epsilon_{s}$is the relative permittivity of the active region. 

This is unpleasant to solve in the general case, but when $\vert eV/kT\vert\ll 1$, we can use the first two terms in a series approximation for the exponential, giving
\begin{eqnarray}\label{pot}
\frac{d^2 V}{dx^2}=\frac{e^2 n_{0}}{\epsilon_{s}\epsilon_{0}kT}V
\end{eqnarray}
Equation (\ref{pot}) has a solution
\begin{eqnarray}\label{V1}
V=V_{0}\exp{(-x/\lambda_{D})}
\end{eqnarray}
where $\lambda_{D}=(kT\epsilon_{s}\epsilon_{0}/e^2 n_{0})^{\frac{1}{2}}$ and is known as the Debye length. Equation (\ref{pot}) shows that a perturbation in potential tends to build up or die away over distances of the order of $\lambda_{D}$.

Major field changes occur over distances greater than $\lambda_{D}$. Further we consider superlattice structure under another boundary conditions taking into account that graphene is quite different from most conventional three-dimensional materials: intrinsic graphene is a semi-metal or zero-gap semiconductor. Put $x=0$ at the surface of the lower graphene  mono-layaer and the potential is assumed to be zero at $x=0$. If the electron gas outside the metal is so rarefied that it can be treated classically. This $V(x)$ increases as $x$ increases from $0$ to infinity. Then $V(\infty)=\infty$, because $n(\infty)=0$, and $V^{'}(\infty)=0$, because the electric field should vanish as $x\rightarrow\infty$. In terms of $V$ Poisson equation can be written as
\begin{eqnarray}\label{Pois}
V^{''}=- \frac{e^2 n_{0}}{\epsilon_{s}\epsilon_{0}} e^{-eV/kT}
\end{eqnarray}
Multiplying the foregoing by $V^{'}$ and integrating it, using the boundary conditions given above, one obtains
\begin{eqnarray}
\frac{1}{2}(V^{'})^2= \frac{n_{0} e^2}{\epsilon_{s}\epsilon_{0}} kT e^{-eV/kT}
\end{eqnarray}
From which it follows that 
\begin{eqnarray}
V^{'}=\left(\frac{2 n_{0} e^2}{\epsilon_{s}\epsilon_{0}} kT\right)^{\frac{1}{2}}e^{-eV/2kT}
\end{eqnarray}
Integrating this result once again, one yields 
\begin{eqnarray}\label{above}
e^{eV/kT}=\left(\frac{2 n_{0}e^2}{\epsilon_{s}\epsilon_{0}kT} \right)^{\frac{1}{2}}(x+\lambda_{D})
\end{eqnarray}
Since we have assumed that $V(0)=0$, on substituting $x=0$ in (\ref{above}), one obtains the value of integrating constant $\lambda_{D}$. Hence (\ref{above}) can be rewritten as $\exp{(eV/kT)}=x/\lambda_{D}+1$, from which
\begin{eqnarray}\label{V2}
V=2V_{0}\log{\frac{x+\lambda_{D}}{\lambda_{D}}}
\end{eqnarray}
On substituting this in (\ref{P}), one finally gets
\begin{eqnarray}
n(x)=n_{0}\left( \frac{\lambda_{D}}{x+\lambda_{D}}\right)^2
\end{eqnarray}
\begin{figure}
	\centering
\includegraphics[scale=0.7]{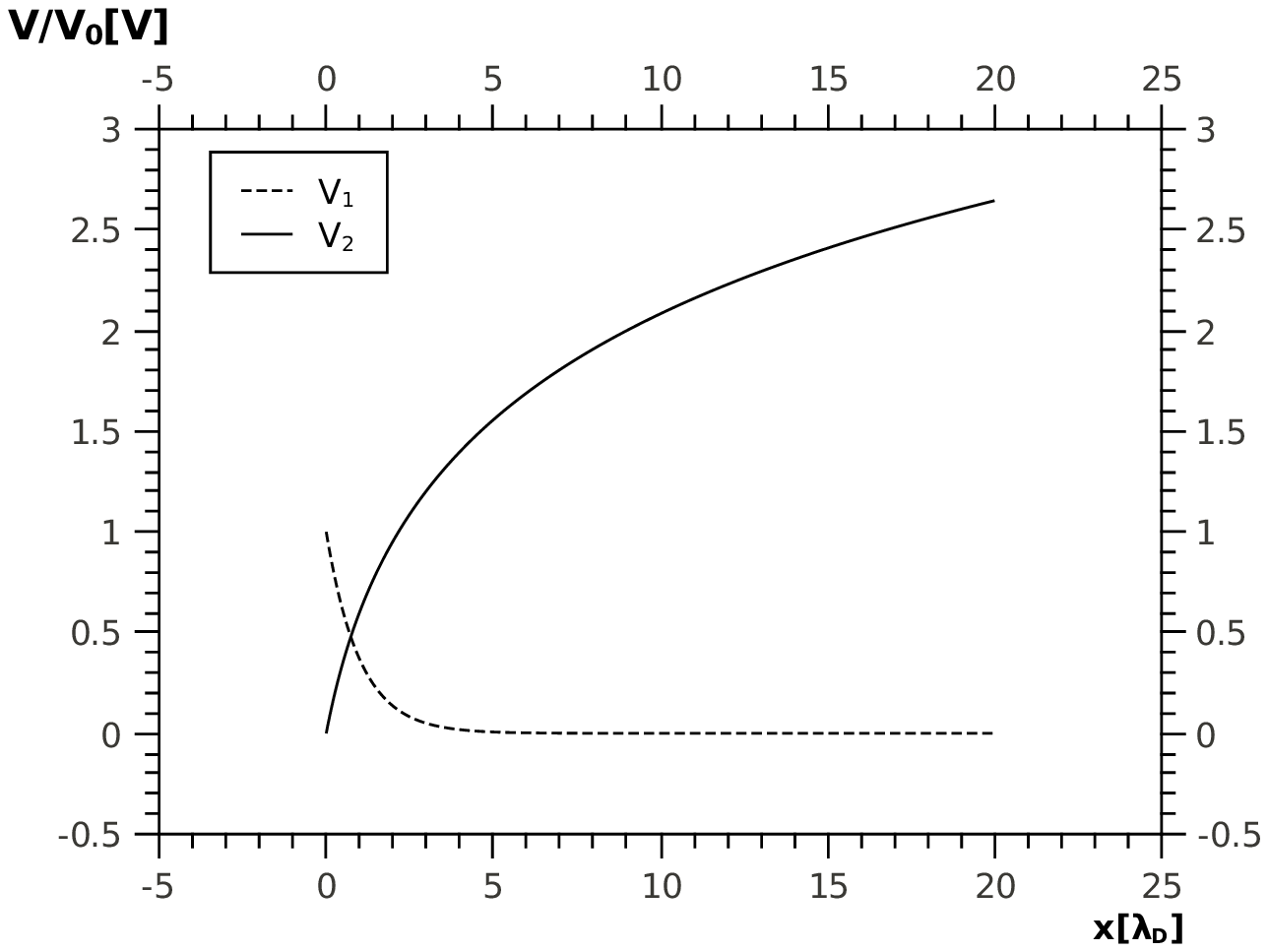} 
 \caption{Electrostatic potential behaviour in term of screening length units. Plotted lines calculated: dashed-$V_{1}$ from (\ref{V1}) and solid-$V_{2}$ from (\ref{V2}) correspondingly. }
	\label{pot}
\end{figure}
\subsection{Boltzmann transport equation}
In the previous section we considered the case of intrinsic semiconductors, the number of electrons which are  excited to the conduction band will be equal to the number of holes in the valence band. The electrical conductivity of electrons or holes in graphene-based superlattice structures due to doping electrons (confined to one material (InAs)) and holes (confined to the other (GaSb)) can be investigated by treating the Boltzman transport equation
\begin{eqnarray}\label{Boltzman}
\frac{\partial f}{\partial t}+ \bold{\upsilon}\cdot\frac{\partial f}{\partial \bold{x}}+\bold{F}\cdot\frac{\partial f}{\partial\bold{k}}=\left( \frac{\partial f}{\partial t}\right)_{coll}
\end{eqnarray}
where $\bold{x}$ is the coordinate, $\bold{k}$ the momentum, $f$ the distribution function of carriers and $\bold{F}$  the external force acting on particle. 
In the paradigm graphene-based modulation-doped superlattice structures of InAs/graphene/GaAs/graphene/GaSb the interface modes in graphene mono-layers emerge as crucial factors and the higher-frequency mode produces symmetric field in the GaAs well that markedly enhance the intrasubband scattering rate.

It is sufficient to find the current density in the form of a term proportional to the electric field. Under the assumptions of steadiness and uniformity, the Boltzmann equation reduces to
\begin{eqnarray}
-e\bold{E}\cdot\frac{\partial f}{\partial\bold{k}}=-\frac{f-f_{0}}{\tau}
\end{eqnarray}
In order to determine an expression correct to first order in $\bold{E}$ (electric field), the distribution function $f$ on the right hand side may be replaced by the zeroth approximation $f_{0}$, i.e. the solution in the case of $\bold{E}=0$. Noticing that $f_{0}$ is a function of $E$, one gains
\begin{eqnarray}\label{fun}
f=f_{0}+\frac{\partial f_{0}}{\partial E}\tau e\upsilon\cdot\bold{E}
\end{eqnarray} 
According to this expression, one sees that electric current is produced by a shift of the center of the Fermi distribution. This is most clearly seen in the case of $\varepsilon(\bold{k})=\hbar\upsilon\bold{k}$: $f\fallingdotseq f_{0}(\varepsilon+\tau e\bold{\upsilon}\cdot\bold{E})$. The current density is obtained by multiplying Eq.(\ref{fun}) by $-e\bold{\upsilon}$ and integrating over all values of the momentum 
\begin{eqnarray}
\bold{j}=-e\int\bold{\upsilon}f\frac{4d\bold{k}}{h^2}=e^2\int\left(-\frac{\partial f_{0}}{\partial\varepsilon}\right)\tau\bold{\upsilon}\bold{\upsilon}\cdot\bold{E}\frac{4d\bold{k}}{h^2}
\end{eqnarray}
Here $d\bold{k}$ stands for $dk_{x}$, $dk_{y}$ and the factor 4 accounts for the weight due to spin and valley. Thus the components of  the electrical conductivity can be written as
\begin{eqnarray}\label{con}
\nonumber
\sigma &=& 4e^2\int\frac{\tau\upsilon^{2}}{3 h^2}\left(-\frac{\partial f_{0}}{\partial\varepsilon}\right)d\bold{k} \\
       &=& \frac{4e^2\tau\bar{\upsilon}^2}{3 k T}\int^{\infty}_{E_{g}}D(E)f(E)dE
\end{eqnarray}
For the Maxwell-Boltzmann distribution, the identity: $-\frac{\partial f_{0}}{\partial \varepsilon}=\frac{1}{kT}f_{0} (f_{0}=e^{(\mu-\varepsilon)/kT})$ holds, and the electrical conductivity is approximate on screening length scale to
\begin{eqnarray}
\sigma=\frac{4 e^2}{3 \pi  h^2}\tau k T e^{-E_{g}/2 k T}
\end{eqnarray}
\begin{figure}
	\centering
	\includegraphics[scale=0.7]{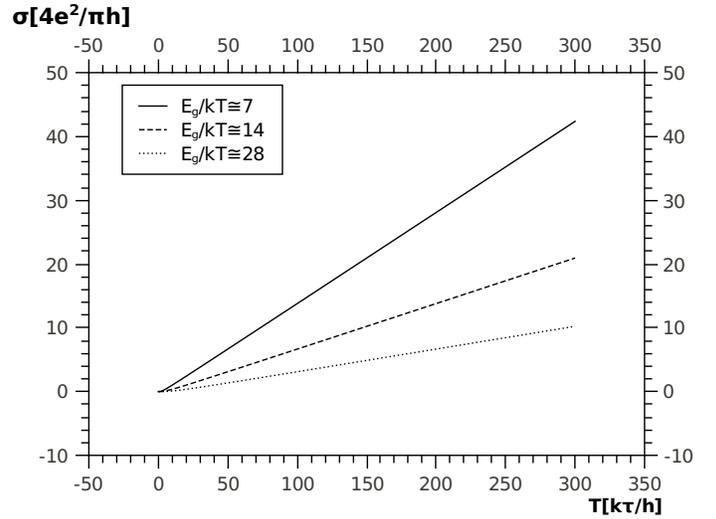} 
\caption{Electrical conductance behaviour of graphene-based superlattice structure shows enhanced mobilities at elevated temperature. Evenly increasing energy gap to thermal energy ($E_{g}/kT\approx 7$, $E_{g}/kT\approx 14$, $E_{g}/kT\approx 28$) tend to steady tunable electrical control and optical confinement on the length scale over screening one.}
	\label{fig-2}
\end{figure}

\begin{figure}
	\centering
	\includegraphics[scale=0.7]{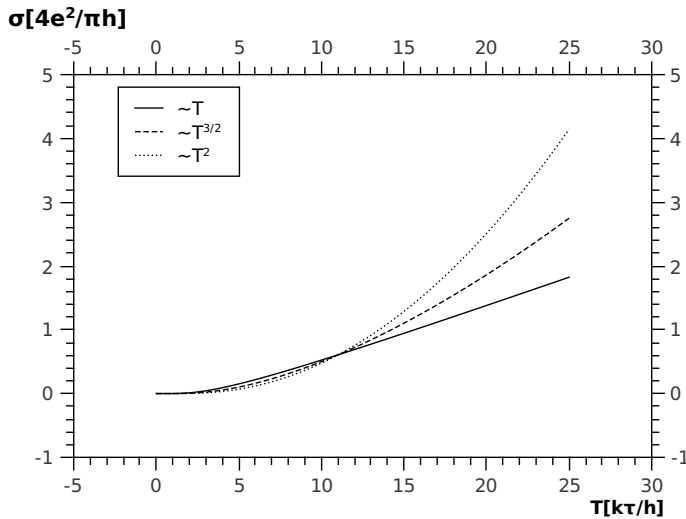}  
\caption{Electrical conductance behaviour (temperature power-law) for $intrinsic$ undoped graphene/GaAs/graphene ($T^2$-like behaviour), undoped GaAs ($T^{\frac{3}{2}}$-like behaviour) and for graphene-based modulation-doped superlattice structure InAs/graphene/GaAs/graphene/GaSb ($T$-like behaviour) respectively. Undoped sample illustrates relatively high conductance unlike doped more steady and less sensitive to temperature  which more valuable for tunable wide-band gap quantum devices.}
	\label{fig-3}
\end{figure}
Electrical conductance behaviour of graphene-based superlattice structures are illustrated on Fig.2 and Fig.3 respectively. On Fig.2 electrical conductance shows enhanced mobilities at elevated temperature increasing energy gap to thermal energy ($E_{g}/kT\approx 7$, $E_{g}/kT\approx 14$, $E_{g}/kT\approx 28$) which tend to steady tunable electrical control and optical confinement on the length scale over screening one.
On Fig.3  displays electrical conductance behaviour (temperature
power-law) for $intrinsic$ undoped graphene/GaAs/graphene structure, undoped GaAs and for graphene-based modulation-doped superlattice structure InAs/graphene/GaAs/graphene/GaSb 
respectively. Undoped sample illustrates relatively high conductance unlike doped more steady and less sensitive to temperature  which more valuable and significant for tunable wide-band gap quantum devices.
\subsection{Thermionic current}
At $0^{\circ}$ K, electrons take the configuration of minimum energy. The electrons in the donors fall into the acceptor levels until the acceptors are all filled, in this configuration the Fermi level must lie at the donor level: $\mu(0^{\circ} K)=\frac{1}{2}E_{g}-E_{d}$. At sufficiently high temperature, the electrons in the filled band can be excited to the conduction band. When the density of the holes in the filled band and the density of the electrons in the conduction band become much larger than the number of donors and acceptors. The effects of donors and acceptors can be neglected, and the sample shows characteristics similar to those of an intrinsic semiconductor. In this case the Fermi level comes in the middle of the energy gap $E_{g}$ and one has: $\mu(\infty)=0$. At the temperatures between these extreme cases ($T=0$ and $T=\infty$), $\mu$ has a value between those given above. Summarizing, one can say that the behaviour of $\mu$ is as follows: at $0^{\circ}$ K, $\mu$ coincides with donor level. It increase with temperature and then it approaches the middle of the gap between the conduction band and the filled band. This is not exact, but is sufficient for a qualitative discussion on the screening length scale. 
\begin{figure}
	\centering
	\includegraphics[scale=0.7]{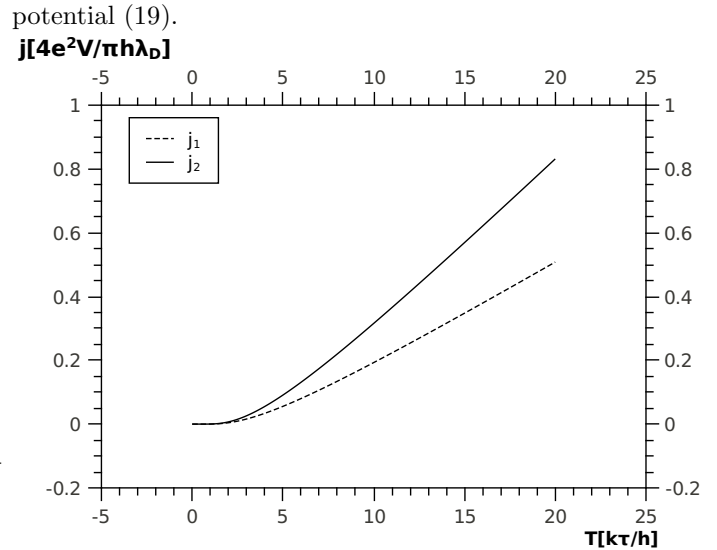} 
\caption{Thermionic current density for different electrostatic potential ( dashed-$j_{1}$ for (\ref{V1}) and solid-$j_{2}$ for (\ref{V2}) potentials correspondingly) fixed on the screening length scale shows temperature dependence.}
	\label{cur}
\end{figure}

At finite temperature electrons having higher energies than work function $W=eV$ at upper tail of the Fermi distribution can escape from graphene surface to the interior of superlattice structure in the direction normal to the surface. When an appropriate potential difference is applied, it is possible to collect all of the electrons escaping from the metal (graphene). For a  graphene-based modulation-doped superlattice structure the electric current density which occurs without any fluctuations in equilibrium can be written as: $\bold{j}=\sigma \bold{E}$ where $\bold{E}$ can be represented as $\bold{E}=V(\bold{x})/\bold{x}$. On the fig.\ref{cur} is illustrated the thermionic current density behaviour which was enhanced due to more realistic intrinsic electrostatic potential (\ref{V2}).
\section{Conclusion}
In this Letter we investigated electric transport properties for a  graphene-based  modulation-doped superlattice structures providing qualitatively good description. We have shown that slightly doped superlattice structures, which based on graphene mono-layers as a high-conductivity channels, in tuning to the point of $intrinsic$-type structures carrier concentrations  can relatively insensitive to heat, generate less noise when operated at high frequencies avoiding scattering effects on screening length scale. The thermionic current density behaviour is enhanced due to more realistic intrinsic electrostatic potential, which was calculated taking into account the effect of metallic (graphene) mono-layers. The proposed structure might become of great use for new types of wide-band energy gap quantum devices. 
\section{ACKNOWLEDGMENT}
Authors are very indebted to Prof.  J. Kwo and Prof. M. Hong for stimulating discussions and fruitful suggestions. We acknowledge the National Center for Theoretical Sciences and Myerscough Bequest for financial support.

\end{document}